\documentclass[sigconf]{acmart}  

\copyrightyear{2025}
\acmYear{2025}
\setcopyright{acmlicensed}
\acmConference[ICMR '25] {Proceedings of the 48th International ACM ICMR Conference on Research and Development in Information Retrieval}{ June 30--July 3, 2025}{Chicago, IL, USA.}
\acmBooktitle{Proceedings of the 48th International ACM ICMR Conference on Research and Development in Information Retrieval (ICMR '25), June 30--July 3, 2025, Chicago, IL, USA}
\acmISBN{979-8-4007-1592-1/25/07}
\acmDOI{10.1145/3731715.3733392}
\usepackage{algorithm}
\usepackage{algorithmic}
\usepackage{amsmath}
\usepackage{amsthm} 
\usepackage{graphicx}
\usepackage{amsfonts}
\usepackage{bm}
\usepackage{adjustbox}
\usepackage{subcaption}
\usepackage{xcolor}
\usepackage{tabularx}
\usepackage{booktabs}  
\usepackage{hyperref}
\usepackage{balance}

\newcommand{\sys}{\textsc{MoE-NP}~}

\newtheorem{theorem}{Theorem} 

\AtBeginDocument{%  
  }

\begin{document}  

%%------------------------------------------------------  
\author{Yu Shi}  
\authornote{Equal contribution.}          % 给 Yu Shi 添加“共同一作”  
\affiliation{%  
  \institution{National University of Defense Technology}  
  \city{Changsha}  
  \country{China}  
}  
\email{shiyu19@nudt.edu.cn}  

\author{Yiqi Wang}  
\authornotemark[1]                        % 复用上一条作者注释（即序号 1）的标号  
\authornote{Corresponding author.}        % 加“通信作者”的注释  
\affiliation{%  
  \institution{National University of Defense Technology}  
  \city{Changsha}  
  \country{China}  
}  
\email{yiq@nudt.edu.cn}  

\author{WeiXuan Liang}  
\affiliation{%  
  \institution{National University of Defense Technology}  
  \city{Changsha}  
  \country{China}  
}  
\email{weixuanliang@nudt.edu.cn}  

\author{Jiaxin Zhang}  
\affiliation{%  
  \institution{National University of Defense Technology}  
  \city{Changsha}  
  \country{China}  
}  
\email{zhangjiaxin18@nudt.edu.cn}  

\author{Pan Dong}  
\affiliation{%  
  \institution{National University of Defense Technology}  
  \city{Changsha}  
  \country{China}  
}  
\email{dongpan@nudt.edu.cn}  

\author{Aiping Li}  
                      % 复用 Pan Dong 的注释（即序号 2）  
\affiliation{%  
  \institution{National University of Defense Technology}  
  \city{Changsha}  
  \country{China}  
}  
\email{liaiping@nudt.edu.cn}

\title{Mixture of Experts for Node Classification}  

\renewcommand{\shortauthors}{Yu Shi et al.}

%%------------------------------------------------------  
%% 以上部分完成了双盲评审需要的信息隐藏配置  
%%------------------------------------------------------  

%%  
%% Start of the document.  

\begin{abstract}
Nodes in the real-world graphs exhibit diverse patterns in numerous aspects, such as degree and homophily. 
However, most existent node predictors fail to capture a wide range of node patterns or to make predictions based on distinct node patterns, resulting in unsatisfactory classification performance. 
In this paper, we reveal that different node predictors are good at handling nodes with specific patterns and only apply one node predictor uniformly could lead to suboptimal result. 
To mitigate this gap, we propose a mixture of experts framework, \sys, for node classification.
Specifically, \sys combines a mixture of node predictors and  strategically selects models based on node patterns. 
Experimental results from a range of real-world datasets demonstrate significant performance improvements from \sys.

\end{abstract}




\maketitle

\section{Introduction}

Node classification, which aims to predict the class of nodes in a graph, has a wide range of applications~\cite{tang2016node,xiao2022graph}, such as citation networks and co-author networks ~\cite{lin2020structure,xiao2022graph}. The key to node classification tasks lies in node representation learning ~\cite{Wei2021Graph}, which has been theoretically and empirically proven to be one of the key strengths of Graph Neural Networks (GNNs)~\cite{hamilton2017representation}. GNNs typically refine node representations through information aggregation and feature transformation among the nodes, and have achieved impressive performance in various graph-related tasks, such as social analysis~\cite{ben2024enhancing}, recommendation system~\cite{wu2022graph} and traffic prediction~\cite{yu2018spatio}.

Most existent GNN models are designed based on the homophily assumption~\cite{bi2024make,zheng2024missing,platonov2024characterizing}, i.e., nodes tend to be similar with those connected to them. However, this assumption is not always true. There exist numerous heterophilic graphs in reality, such as the social network in a dating website~\cite{wang2024understanding}, where numerous users tend to follow people of different genders. To solve node classification in these scenarios, several studies propose to integrate information from broader neighborhoods so as to strengthen the relationships between originally-unconnected nodes. For instance, MixHop~\cite{abu_Mixhop_2019} repeatedly mixes feature representations of neighbors at various distances. Geom-GCN~\cite{pei2020geom} proposes to add virtual nodes in feature aggregation. WRGCN~\cite{suresh2021breaking} constructs a modified computation graph to enhance the connections and information for disassortative nodes. Furthermore, some recent work~\cite{luan_revisiting_2022,Chen_LSGNN_2023} proposes to deal with this heterophliy challenge in a node-wise manner, given the consideration that different nodes in the same graph can be faced with distinct heterophliy challenges.

\begin{figure*}[!ht]
\begin{subfigure}{0.49\textwidth}  
    \centering  
    \includegraphics[width=\textwidth]{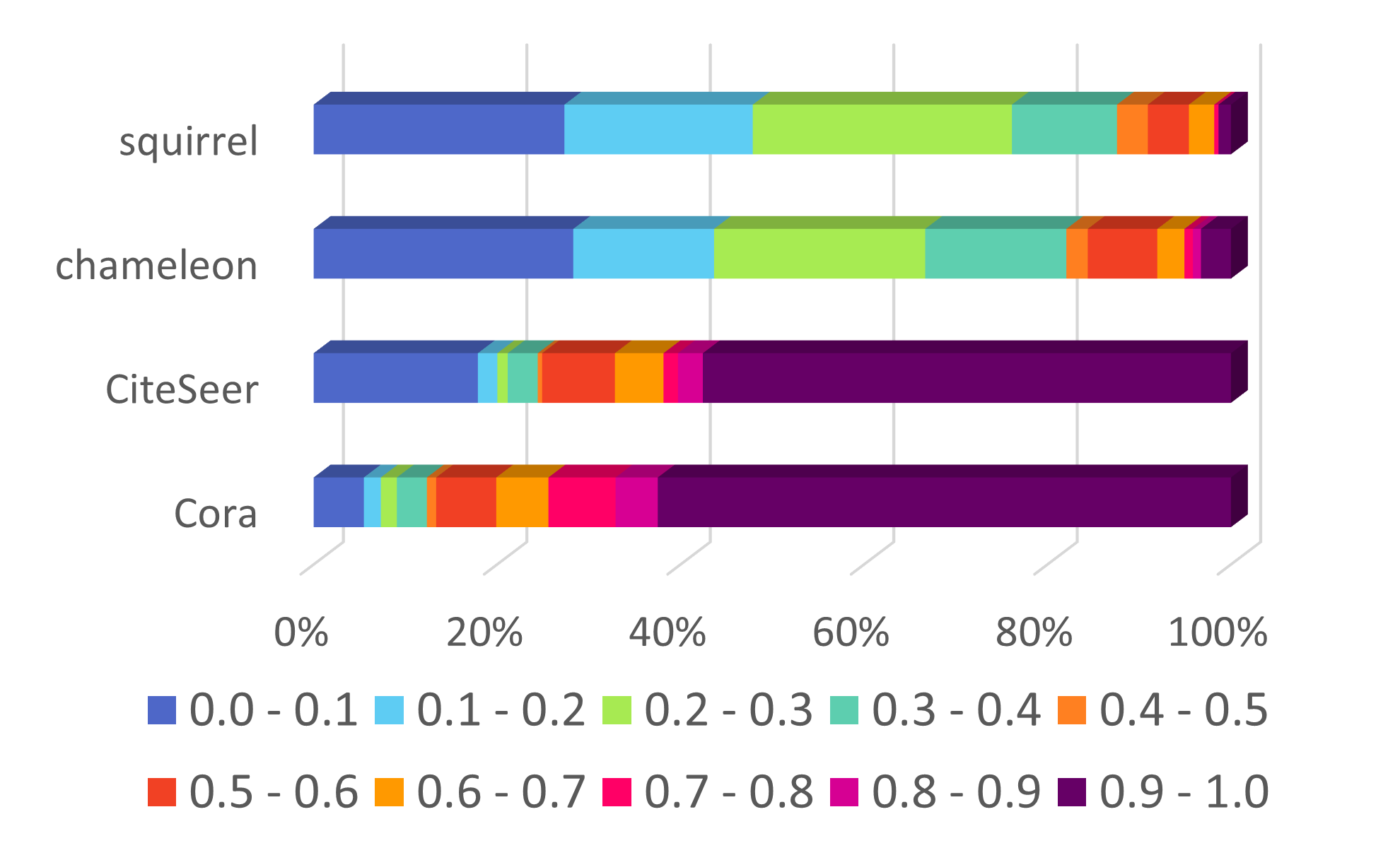}
    \caption{homophily diversity}
    \label{fig1_a}  
\end{subfigure}  
\hfill  
\begin{subfigure}{0.49\textwidth}  
    \centering  
    \includegraphics[width=\textwidth]{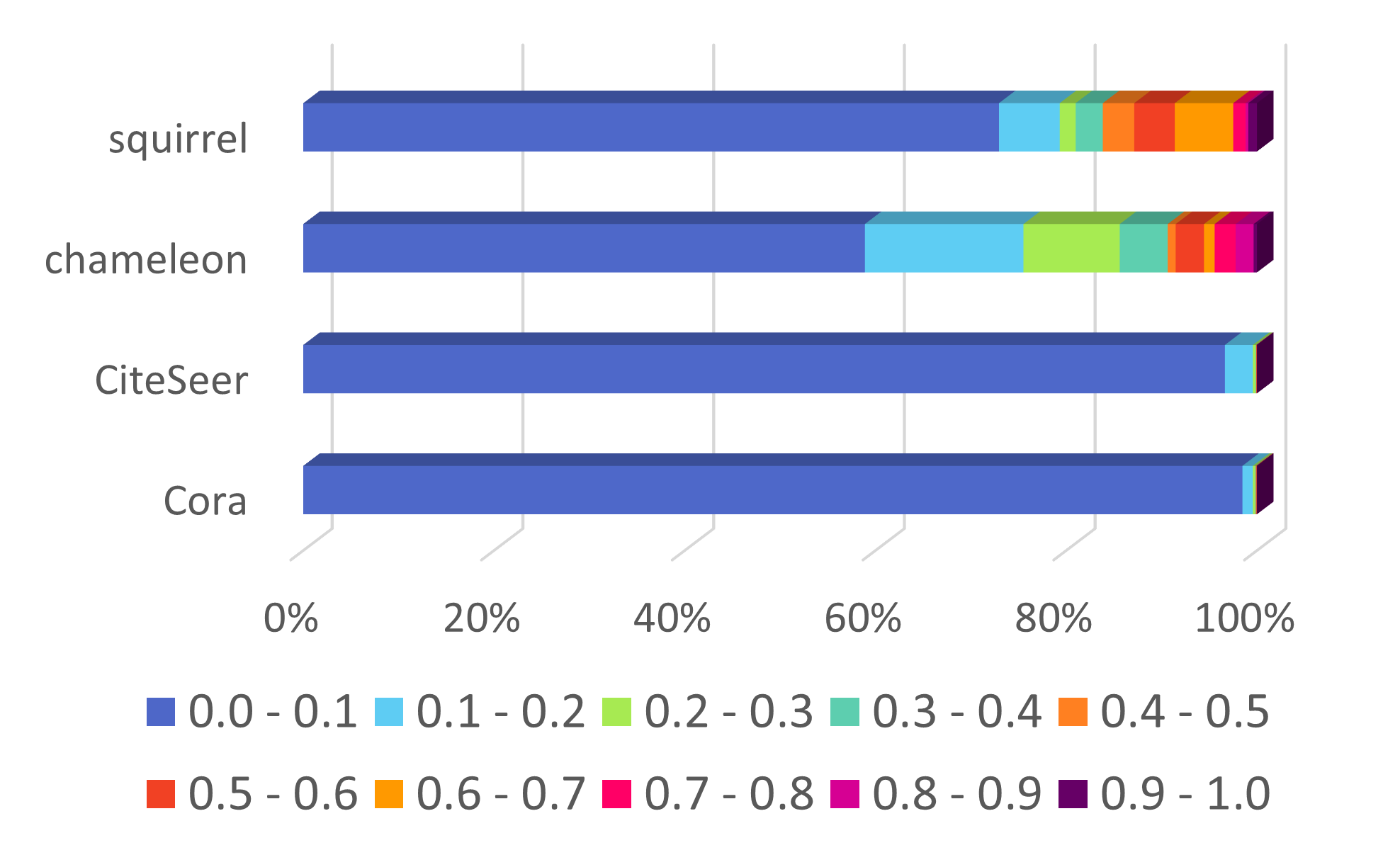}
    \caption{degree diversity}
    \label{fig1_b}  
\end{subfigure}  
\caption{The distribution of node homophily and degree across different datasets}
\label{fig1}
\end{figure*}

While heterophily issue has been widely studied, some latest research~\cite{ma2021homophily,mao2024demystifying} suggests that node classification is essentially related to the node pattern, rather than merely the label homophliy. However, the node pattern is quite complex, which can be described from various perspectives, including label homophily, structure information and attribute information. In addition, nodes within the same graph can possess diverse node patterns from different perspectives. This naturally raises a problem:\textit{ Do nodes with different node patterns within the same graph require distinct node predictors?}

To answer this question, we first conducted some empirical investigations over node patterns. We have observed that nodes with different patterns in one graph require distinct node classifiers and their preferences for classifiers are affected by numerous factors. Furthermore, we have noticed that the classifier preference is not only related to node pattern, but also associated with the overall graph context. Based on these observation, we then propose~\sys, a mixture-of-expert model for the node classification task, which consists of a bunch of diverse node predictors and a specially designed gating network. The gating network is able to adaptively assign experts to each node according to both its local node pattern and the overall graph context. Extensive experiments have shown the superior performance of the 
proposed~\sys. Specifically, it surpasses the best baseline on PubMed  and Actor datasets by 4.4\% and 3.4 \% in accuracy. In addition, a theoretical analysis is provided to demonstrate the rationality of~\sys.

\section{Preliminary}

In this section, We investigate node patterns within the same graph from different perspectives and their influence towards the classifiers. Some notations and backgrounds are first introduced, and then we discuss the empirical observations and key insights.

\begin{figure*}[!ht]
\begin{subfigure}{0.32\textwidth}  
    \centering  
    \includegraphics[width=\textwidth]{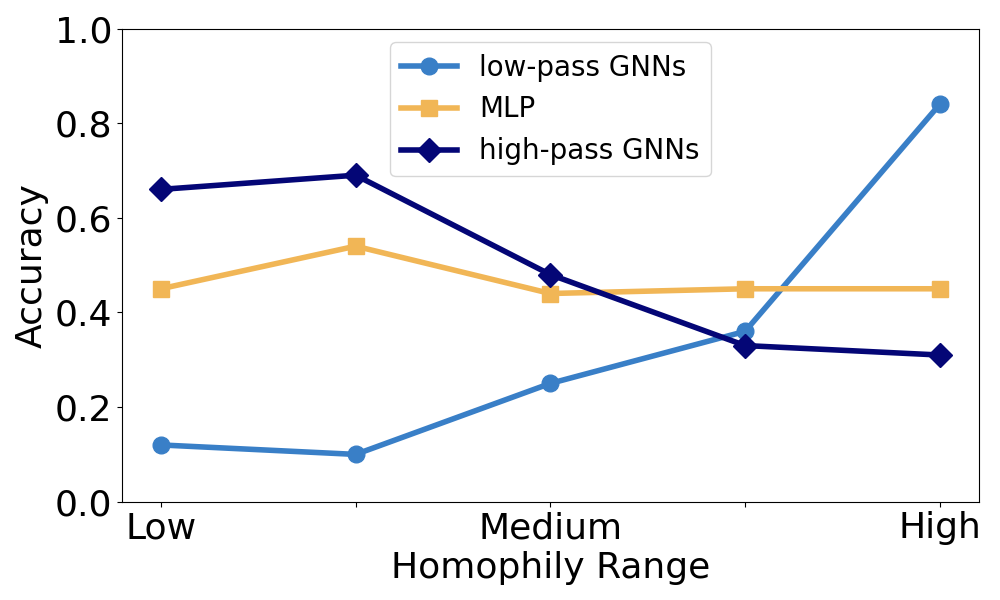}   
    \caption{Chameleon(Homophily)}  
    \label{fig2_a}  
\end{subfigure}  
\hfill  
\begin{subfigure}{0.32\textwidth}  
    \centering  
    \includegraphics[width=\textwidth]{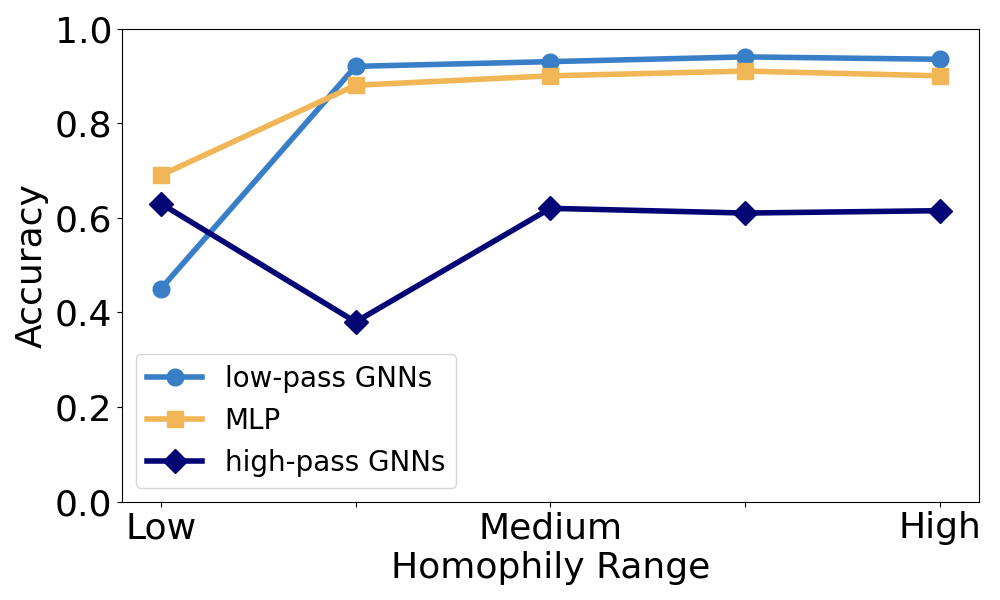}  
    \caption{PubMed(Homophily)} 
    \label{fig2_b}  
\end{subfigure}
\begin{subfigure}{0.32\textwidth}  
    \centering  
    \includegraphics[width=\textwidth]{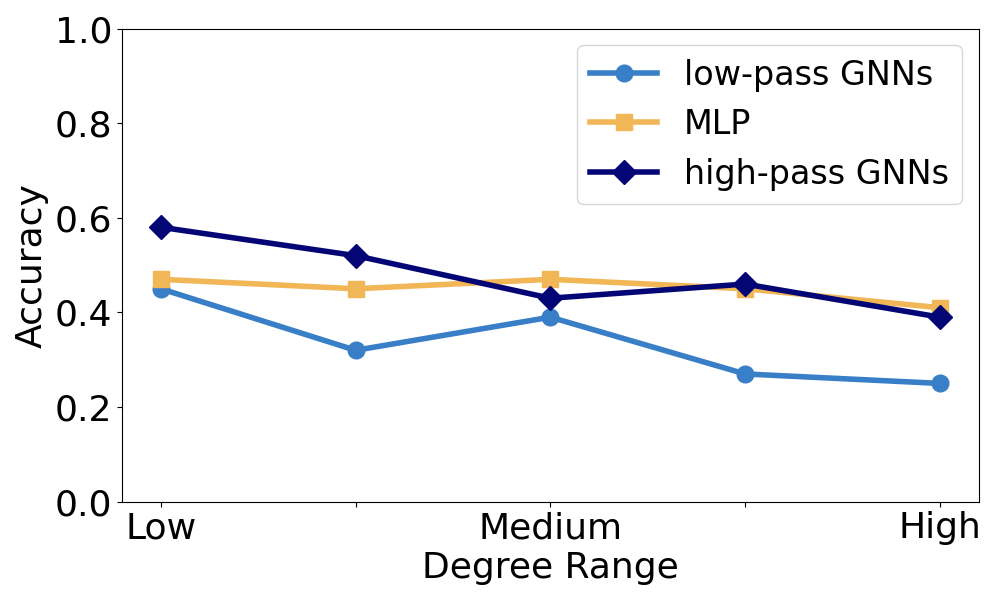}  
    \caption{Chameleon(Degree)}  
    \label{fig2_c}  
\end{subfigure}  

\caption{Figure(a) and (b)  show the classifier preferences for nodes from different homophily groups  in Chamelon and PubMed, where Chamelon is a typical heterophlic graph and PubMed is a homophlic one. Figure(c) further demonstrates the classificaiton performance of different node classifiers in node groups with various degree values from the low homophily group in Figure(a).}
 \vspace{-0.3in}
\label{fig2}
\vspace{+0.1 in}
\end{figure*}

\subsection{Notations and Backgrouds}
We denote an undirected graph without self-loops as $\mathcal{G} =  (\mathcal{V},\mathcal{E})$, where $\mathcal{V} = \{v_i\}_{i=1}^n$ is the node set and $\mathcal{E}\subseteq\mathcal{V}\times\mathcal{V}$ is the edge set. Let $\mathcal{N}\left(v_i\right)$ denote the neighbors of node $v_i$. $\mathbf{A}\in \mathbb{R}^{n\times n}$ denotes the adjacency matrix, and $\mathbf{D}$ represents the diagonal matrix standing for the degree matrix such that $\mathbf{D}_{ii}=\sum_{j=1}^n\mathbf{A}_{ij}$. Let $\tilde{\mathbf{A}}$ and $\tilde{\mathbf{D}}$ be the corresponding matrices with self-loops, i.e., $ \tilde{\mathbf{A}}=\mathbf{A}+\mathbf{I}$ and $\tilde{\mathbf{D}}=\mathbf{D}+\mathbf{I}$, where $\mathbf{I}$ is the identity matrix. 
% $\mathbf{X} = \{x_i\}_{i=1}^n \in \mathbb{R}^{n\times d}$ 
$\mathbf{X} \in \mathbb{R}^{n\times d}$ 
denotes the initial node feature matrix, where $d$ is the number of dimensions.
We use $y$ to denote the ground-truth node label and $y_i$ denote the label of node $i$. In this paper, we  define homophily $h$ on graph $\mathcal{G}$  as 

\begin{align}
    h(\mathcal{G})=\frac{1}{|\mathcal{V}|}\sum_{v_i\in\mathcal{V}}\frac{\left|\{u\mid u\in\mathcal{N}_{\left(v_i\right)},y_{u}=y_{v_i}\}\right|}{\mathbf{D}_{ii}}.
\end{align}
where $h(\mathcal{G})$ is the average node homophily value for Graph $\mathcal{G}$, within the range of $[0,1]$. 

Most Graph Neural Networks (GNNs) are designed to use low-pass filters to refine node representation, which tend to smooth node features across the graph based on the homophily assumption. On the contrary, the high-pass GNNs are designed to capture differences between a node and its neighbors, which are more suitable for heterophilic graphs, where nodes tend to connect with different counterparts.
% High-pass GNNs emphasize discrepancies in node information. This makes them effective for graphs with heterotrophic patterns. 
In addition, there exist some models, such as multilayer perceptrons (MLPs), not aggregating neighbor information. Instead, they solely transform features based on the its own data, ignoring the graph structure.

\subsection{Empirical Study}

As illustrated in Figure \ref{fig1}, nodes within the same graph can exhibit significant variability in their patterns when assessed from various perspectives, including homophily and degree. Notably, irrespective of whether the overall graph is characterized as homophilic or heterophilic, the distributions of nodes, in terms of their homophily and degree values, can display substantial diversity within a single graph. Specifically, it is important to recognize that considerable heterogeneity among nodes can exist in graphs generally classified as homophilic, and conversely, graphs deemed heterophilic may also contain nodes manifesting homophilic characteristics.

To investigate the impact of node patterns on the accuracy of node classification, we have conducted a series of preliminary studies. In these studies, nodes from a single dataset were categorized into five distinct groups based on their homophily values, and we subsequently analyzed their affinities for various node classifiers. The findings pertaining to both homophilic and heterophilic graphs are presented in Figure \ref{fig2}. It is evident that nodes belonging to different homophily groups demonstrate diverse preferences for specific node classifiers. Furthermore, this variation in preferences is not solely attributable to the homophily group itself; rather, it is influenced by the overall homophilic nature of the graph. This is corroborated by observations indicating that the most effective classifiers for nodes in the low and medium homophily groups differ significantly between the Chameleon and PubMed datasets.

To further elucidate the diversity of node patterns and their implications for node classification, we conducted a more granular analysis of one homophily group, dividing it into five distinct subsets based on node degree values. As demonstrated in Figure \ref{fig2_c}, even within the same homophily group, the degree of individual nodes markedly influences the performance and effectiveness of various node predictors.

The observed variations in classifier effectiveness across different homophily and degree groups suggest that complex node patterns play a critical role in node classification tasks. Consequently, it is essential to represent node patterns from multiple perspectives. Moreover, both local properties and the broader structural context of the graph significantly affect preferences regarding node classifiers. These insights underscore the necessity of assigning distinct classifiers to different nodes based on their unique patterns across various dimensions, as well as considering the overarching context of the graph itself.

\subsection{Theoretical analysis}

In this section, we provide a theoretical foundation to show that a single node predictor may be inadequate for nodes exhibiting distinct connectivity patterns. For simplicity, the derivation is presented in the context of a binary classification problem. Our proof builds upon the conclusions of \cite{Baranwal2021GraphCF} and \cite{han2024node}, further extending their results.

Our analysis is based on a widely used graph generative model, the Contextual Stochastic Block Model (CSBM) \cite{deshpande2018contextual}. CSBMs assume that edges are formed in either a homophilic or heterophilic manner. Nodes with the same label connect with a probability of \(p\). In contrast, nodes with different labels connect with a probability of \(q\).

The training subgraph is generated from
\begin{align}
        (\mathbf{A}, \mathbf{X}) \sim \operatorname{CSBM}(n, p, q, \boldsymbol{\mu}, \boldsymbol{\nu}), \label{ineq.2}
\end{align}
where \(n\) is the number of training nodes. The vector \(\boldsymbol{\mu}\) and \(\boldsymbol{\nu}\) denote the average feature vectors for  different classes. 
We assume that \(\|\boldsymbol{\mu}\|,\|\boldsymbol{\nu}\|\le 1\). The node features are sampled from a Gaussian distribution \(\mathcal{N}(\boldsymbol{\mu},\sigma^2 \mathbf{I})\), where \(\mathbf{I}\) is the identity matrix.

Test the model on another subgraph generated from a distinct CSBM:
\begin{align}
    (\mathbf{A}^{\prime}, \mathbf{X}^{\prime}) \sim \operatorname{CSBM}(n, p^{\prime}, q^{\prime}, \boldsymbol{\mu}, \boldsymbol{\nu}), \label{ineq.3}
\end{align}
where \(n^{\prime}\) denotes the number of test nodes, and \(p^{\prime}\) and \(q^{\prime}\) represent the intra-class and inter-class edge probabilities, respectively.

A Simplified Graph Convolution (SGC) is employed to update the node embeddings, defined as follows:
\begin{align}
    \tilde{\mathbf{X}} = \mathbf{D}^{-1}\mathbf{A}\mathbf{X}. \label{ineq.4}
\end{align}

A linear classifier parameterized by \(\mathbf{w} \in \mathbb{R}^{d\times1}\) and \(b \in \mathbb{R}\) predicts labels as follows:
\begin{align}
\mathbf{\hat{y}}=\sigma(\tilde{\mathbf{X}} \mathbf{w} + b \mathbf{1}),\label{ineq.5}
\end{align}
where \(\sigma(x)=\left(1+e^{-x}\right)^{-1}\) represents the sigmoid function.

The learning objective is defined by the binary cross-entropy loss function:
\begin{align}
L(\mathcal{V}, \mathbf{w}, b)
= -\frac{1}{|\mathcal{V}|} \sum_{i \in \mathcal{V}}
\bigl[\, \mathbf{y}_i \log \bigl(\hat{\mathbf{y}}_i\bigr)
+ (1-\mathbf{y}_i) \log \bigl(1-\hat{\mathbf{y}}_i\bigr) \bigr],
\label{ineq.6}
\end{align}
where \(\mathcal{V}\) denotes the set of examples used for evaluation.

To ensure linear separability, additional constraints must be imposed by requiring the distance between the class means \(\boldsymbol{\mu}\) and \(\boldsymbol{\nu}\) to satisfy a specific asymptotic lower bound. Formally, the separation condition \(\|\boldsymbol{\nu} - \boldsymbol{\mu}\| = \Omega\left(\tfrac{\log n}{d_n(p+q)}\right)\)  and \(||\mathbf{w}|| \leq R\) must hold, where \(\Omega(\cdot)\) denotes the asymptotic lower bound complexity notation, \(d_n\) represents the dimension-dependent coefficient, and \(p,q\) correspond to the model parameters in the given context.

Furthermore, a sufficiently large \(n\) (specifically, \(\omega(d \log d)\), while remaining \(\mathrm{poly}(d)\)) is required. Moreover, the graph should not be excessively sparse. This stipulation ensures that

\begin{align}
p, q, p^{\prime}, q^{\prime} = \omega\left(\tfrac{\log^2(n)}{n}\right).
\label{ineq.7}
\end{align}

Based on the findings in \cite{Baranwal2021GraphCF}, it has been established that, under specific separability conditions, the optimal linear classifier for the training set \((\mathbf{A}, \mathbf{X})\) takes the form:
\begin{align}
\mathbf{w}^* = R \frac{\boldsymbol{\nu} - \boldsymbol{\mu}}{\|\boldsymbol{\nu} - \boldsymbol{\mu}\|}, 
\quad
\mathbf{b}^* = -\tfrac{1}{2}\bigl\langle\boldsymbol{\nu} + \boldsymbol{\mu}, \mathbf{w}^*\bigr\rangle.
\label{ineq.8}
\end{align}

Next, analyze the performance of the optimal classifier \(\mathbf{w}^*, b^*\) on the new node pattern graph \(\bigl(\mathbf{A}^{\prime}, \mathbf{X}^{\prime}\bigr)\). Our aim is to demonstrate that a single linear classifier may not effectively generalize across different connection patterns determined by \((p, q)\) and \((p', q')\).

\begin{theorem}
\label{Theorem 1}
If \((p - q)(p^{\prime} - q^{\prime}) \leq 0\), with high probability, the following inequality holds:
\begin{align}
    L\left(\mathbf{A}^{\prime},\mathbf{X}^{\prime}, \mathbf{w}^*, b^*\right) 
    \geq \frac{R\left(q^{\prime}-p^{\prime}\right)}{2\left(p^{\prime}+q^{\prime}\right)}\|\boldsymbol{\mu}-\boldsymbol{\nu}\|(1+o(1)),
\end{align}
where 
\(\mathbf{w}^*\), \(b^*\) 
minimize 
\(L\left(\mathbf{A},\mathbf{X}, \mathbf{w}, b\right)\). 
That is, the generalization loss becomes significantly large when the signs of \(p-q\) and \(p'-q'\) differ.

\end{theorem}

\begin{theorem}
\label{Theorem 2}
If the sample sizes of the two categories are equal, \((p - q)(p' - q') \geq 0\), and \(p + q \neq p' + q'\), then the following inequality holds:
\begin{align}
    L\left(\mathbf{A}^{\prime},\mathbf{X}^{\prime}, \mathbf{w}^*, b^*\right) \geq \log(2) \left(1 - \frac{R \|\boldsymbol{\mu} - \boldsymbol{\nu}\| |p' - q'|}{\sqrt{8}\sigma (p' + q')}\right)(1 + o(1)).  
\end{align}
where \(\mathbf{w}^*\), \(b^*\) 
minimize  \(L\left(\mathbf{A},\mathbf{X}, \mathbf{w}, b\right)\). That is, when \(p - q\) and \(p' - q'\) are the same, but \(p + q \neq p' + q'\), i.e., the degree distributions of two sets are different, the generalization loss is also large.
\end{theorem}

\noindent
\textbf{Remark.}
Theorem~\ref{Theorem 1} demonstrates that when the homophily assumption holds for graphs in the training set while not holding for those in the test set (i.e., \((p-q)\) and \((p'-q')\) possess opposite signs), a classifier trained solely on \(\bigl(\mathbf{A},\mathbf{X}\bigr)\) incurs a large loss on \(\bigl(\mathbf{A}^{\prime},\mathbf{X}^{\prime}\bigr)\). Theorem~\ref{Theorem 2} further indicates that even when the homophily conditions are identical in two sets, the performance of a single-node predictor can still be negatively affected by degree disparity. More details can be found in the appendix \ref{proof}.

\section{Method}
\label{Section: Method}
\begin{figure*}[!ht]
\centering
\includegraphics[width=2\columnwidth]{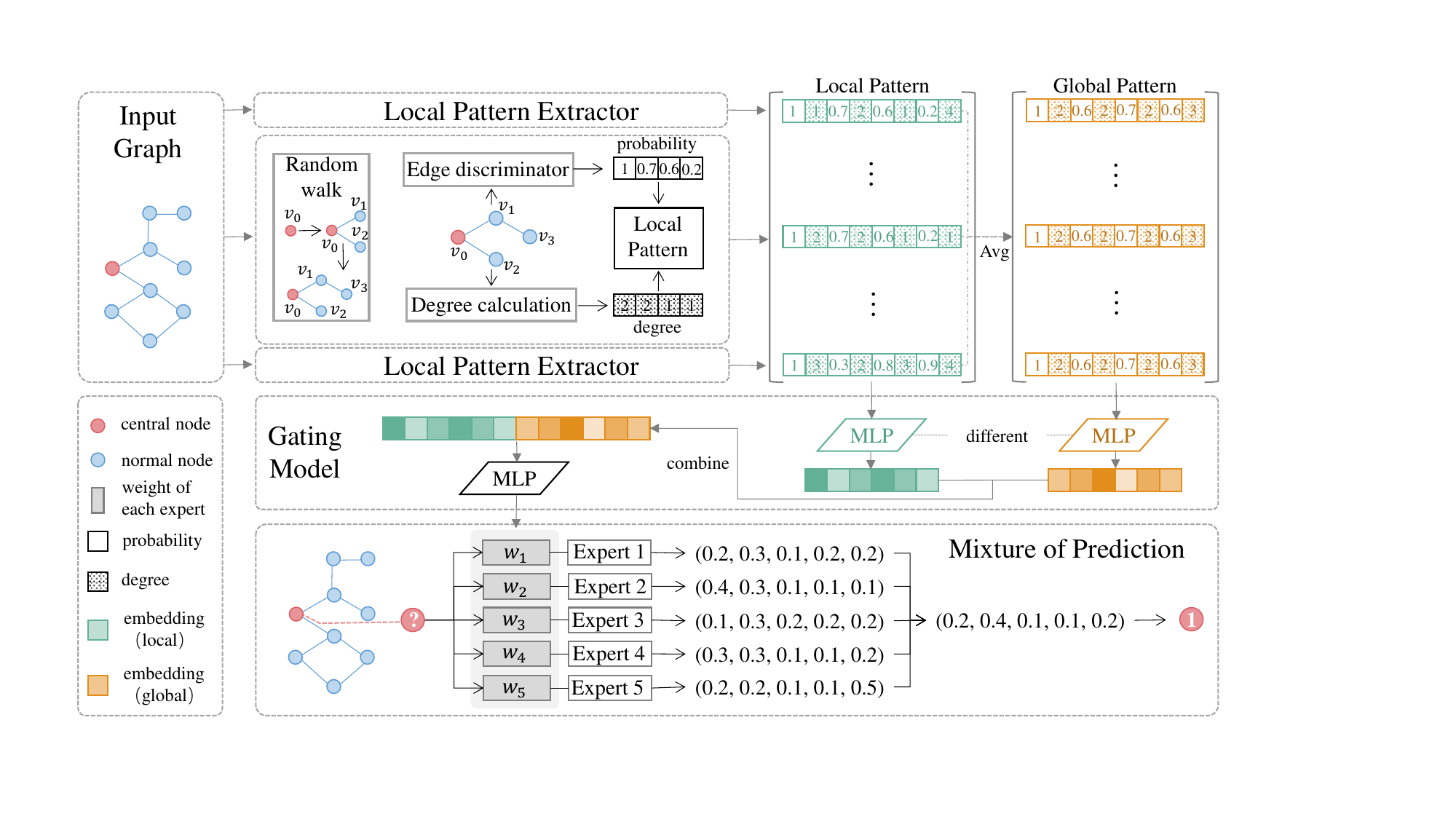} 
\caption{The overall framework of the proposed \sys. For each node, the gating model will assign different weights for each expert based on the node’s pattern.}
\label{fig3}
\end{figure*}
Based on the preliminary study, nodes within the same graph can exhibit diverse node patterns from different perspectives, and this affects their preference for node predictors significantly. To accommodate the node pattern diversity, this paper proposes a Mixture of Experts (MoE) framework to solve node classification, which is able to assign node predictors with different weights to each node based on its node patterns. There are several challenges to be addressed. First, how to accurately define node patterns given they are complex and multi-modal? Second, how to appropriately combine different node predictors for distinct node? In the following sections, we will give an overall framework description, and then illustrate how we address these challenges.

\subsection{\sys – A General Framework}

The overall  \sys framework is illustrated in Figure \ref{fig3}, which consists of two main components, the node pattern extractor and the gating model. Specifically, the node pattern extractor is designed to extract multi-modal node patterns for each node. The gating model is developed to generate node-wise expert weights for different nodes based on their extracted node patterns. There are five common node predictors adopted as experts in the proposed \sys, and more advanced experts can be easily added into the flexible \sys framework.

\subsection{Node Pattern Extractor}

As highlighted in prior research, the presence of distinct node patterns exerts a significant impact on the accuracy of node classification. Consequently, it is paramount to develop a robust node pattern extractor for the proposed \sys. However, designing such an extractor is nontrivial. First, node patterns encompassing a range of properties such as degree distributions and label homophily distributions. Second, computing label homophily distributions precisely remains challenging because the test graph lacks ground-truth labels.

To address these issues, we propose a carefully structured node pattern extractor. Specifically, we employ random walk sampling in each node’s neighborhood to objectively capture local structural context. The sampled nodes form a local context for the target node, within which pattern extraction is performed. Recognizing that node labels are unavailable, \sys introduces an edge discriminator—implemented as a multi-layer perceptron (MLP)—designed to predict the relationship between a target node and its sampled neighbors based on their node features. By learning this relationship, the system effectively harnesses node feature information to compensate for the absence of explicit label data.

Furthermore, to incorporate structural information more comprehensively, we concatenate the degree of each node with its corresponding edge discriminator output on a node-by-node basis, thereby creating an enriched representation of local patterns. As depicted in Figure~\ref{fig3}, this design offers a systematic approach to extracting node patterns, ultimately improving classification performance in challenging graph settings.

\begin{table*}[t]
\centering  
\begin{adjustbox}{max width=\textwidth}  
\begin{tabular}{c c c c c c c c c}
\toprule  
& \multicolumn{2}{c}{Homophilic Datasets} & \multicolumn{5}{c}{Heterophilic Datasets} & \begin{tabular}{c} Avg \\ Rank \end{tabular} \\
\cmidrule(lr){2-3} \cmidrule(lr){4-8}  models& Cora & PubMed & Texas & Cornell & Wisconsin & Chameleon & Actor & \\
\midrule
MLP & $75.01 \pm 2.29$ & $87.78 \pm 0.3$ & $76.76 \pm 4.55$ & $78.11+8.15$ &$81.37+6.63$  &$49.25+2.12$ &$36.53+0.78$&6.14 \\
GCN & $87.26 \pm 1.24$ & $88.02 \pm 0.49$ & $62.43 \pm 5.05$ & $62.43 \pm 3.3$ & $60.59 \pm 7.82$ & $68.11 \pm 1.32$ & $30.57 \pm 0.74$&6.43 \\
HighPass GCN & $38.15 \pm 2.21$ & $63.74 \pm 0.89$ & $78.11 \pm 4.43$ & $59.73 \pm 5.05$ & $74.51 \pm 5.11$ & $59.50 \pm 2.4$ & $32.71 \pm 1.11$ &6.71 \\
ACMGCN & $88.27 \pm 1.15$ & \underline{$89.74 \pm 0.5$} & $\mathbf{87.30 \pm 2.97}$ & $77.57 \pm 4.84$ & $84.9 \pm 2.78$ & $68.62 \pm 1.8$ & $36.07 \pm 1.29$ & 3.29\\
LINK & $78.91 \pm 2.08$ & $80.51 \pm 0.7$ & $65.41 \pm 4.49$ & $55.68 \pm 7.07$ & $61.57 \pm 6.4$ & $70.99 \pm 2.16$ & $24.26 \pm 1.1$ & 6.57 \\
LSGNN & $87.36 \pm 0.89$ & $89.33 \pm 0.44$ & $81.08 \pm 3.63$ & $79.73 \pm 5.57$ & $84.31 \pm 4.02$ & $\mathbf{74.12 \pm 1.33}$ & $35.50 \pm 1.29$ &3.57 \\
GloGNN & \underline{$88.31 \pm 1.13$} & $89.62 \pm 0.35$ & $84.32 \pm 4.15$ & \underline{$83.51 \pm 4.26$} & $\mathbf{87.06 \pm 3.53}$ & $69.78 \pm 2.42$ & \underline{$37.35 \pm 1.30$} &\underline{2.43} \\
\textbf{\sys}& $\mathbf{89.31 \pm 0.91}$ & $\mathbf{94.02\pm0.33}$ & \underline{$86.76 \pm 4.26$} & $\mathbf{86.49 \pm 5.54}$ & \underline{$85.29 \pm 3.07$} & \underline{$73.82 \pm 2.1$} & $\mathbf{40.74 \pm 1.36}$ &\textbf{1.43}  \\

\bottomrule 
\end{tabular}
\end{adjustbox}

\caption{The results of the node classification experiments on both homophilic and heterophilic datasets}
\label{table1}
\end{table*}

\subsection{Design of the Gating Model}

Building on the local patterns generated by the node pattern extractor, it is crucial to construct a gating model capable of adaptively determining the relative importance of each node classifier in accordance with the node’s specific pattern characteristics. Prior research \cite{mao2024demystifying} and preliminary analyses reveal that node classification depends not only on local node attributes but also on broader graph-wide characteristics. For example, nodes with low homophily may necessitate specialized classifiers due to the variations in the underlying graph.  

To capture such global properties, \sys computes a global graph pattern by averaging the local patterns across all nodes within the graph. Both the local node pattern and the global graph pattern are then independently transformed into a uniform embedding space by means of two learnable multi-layer perceptron (MLP) models. Subsequently, the embeddings from these two perspectives—local and global—are concatenated to construct a more comprehensive representation of each node and the graph it resides in.  

Finally, this concatenated representation is fed into the gating model, which is itself an MLP equipped with a softmax activation function. The softmax outputs are interpreted as weights that reflect the contribution of each expert classifier to the final prediction for a given node. Through this mechanism, \sys effectively leverages both local and global contextual information to optimize the assignment of node classification tasks among the available experts.

\subsection{Optimization Objective}

Because \sys aims primarily at improving node classification, it employs a canonical node classification loss as the overarching optimization objective. Formally, the loss function is defined as:
\begin{equation}
    L_{\mathrm{MoE\_NP}} = \sum_{i=1}^n \log\bigl(1 + \exp\bigl(-y_i \bigl(\sum_{j=1}^t w_j E_j(x)\bigr)\bigr)\bigr),
\end{equation}
where \(n\) is the total number of training nodes, \(y_i\) is the true label of node \(i\), and \(E_j(\cdot)\) represents the prediction output of the \(j\)-th expert. The weights \(w_j\), determined by the gating model, modulate the contribution of each expert’s prediction to the final output. By simultaneously optimizing the edge discriminator and the gating model in accordance with this objective, \sys enables end-to-end learning of both the node pattern extractor and the expert gating model, thus improving classification performance.

\section{Experiments}

In this section, comprehensive experiments have been conducted to validate the effctiveness and rationality of the propsoed ~\sys. The experiments settings are first introduced, then the comparison of \sys and the state-of-the-art baseline on different datasets are demonstrated. Next comes further analysis of the  proposed \sys.

\subsection{Experimental settings.}

\textbf{Datasets.} We have conducted experiments on seven widely-used datasets, including two homophilic datasets, Cora and Pubmed~\cite{sen2008collective} and five heterophilic datasets, Texas, Wisconsin, Cornell, Chameleon,  and Actor~\cite{rozemberczki2021multi}. For each dataset, 10 random splits are generated with 60\% training, 20\% validation, and 20 \% testing partitions. In each split, the models are trained with 3 different random seeds, and the average performance and standard deviation are reported.

\begin{figure}[!ht] 
\centering  
\includegraphics[width=\linewidth]{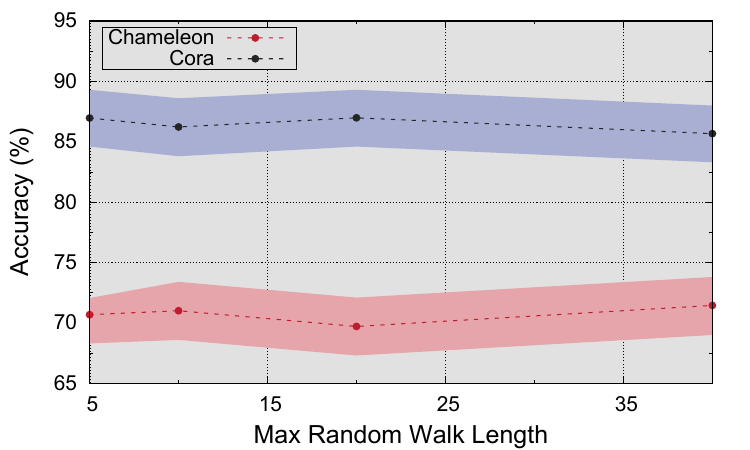}
\caption{Performance on Cora and Chameleon under different random walk lengths (5, 10, 20, 40).}
\label{fig5}  
\end{figure}

\textbf{Baselines.} Seven baselines have been adopted for comparison, including basic methods, such as  MLP, common GCN~\cite{kipf2016semi}and high-pass GCN, and some latest methods that are specially designed for heterotrophic graphs, such as ACMGCN~\cite{luan_revisiting_2022}, GloGNN~\cite{li2022finding}, LinkX~\cite{lim2021linkx} and LSGNN~\cite{Chen_LSGNN_2023}.

\textbf{\sys Implementations.} In this work, we implement five commonly-used models as experts. They are GCN, GCN with residual connect, High-pass GNN, High-pass GNN with residual connect, and MLP. Additional node predictors can be easily added as experts, given the flexibility of \sys.

\subsection{The Overall Performance Comparison}
We have compared the proposed \sys with seven baselines on seven datasets, including both the homophilic and heterophilic graphs. The results are shown in Table \ref{table1}. The proposed \sys has demonstrated impressive performance on both the homophilic and heterophilic graphs, achieving the best overall rank among all the methods. The basic baselines, such as MLP and GCN, tend to perform well only on specific groups, while \sys leveraging several such baselines as experts can achieve significant improvements in terms of classification performance, even surpassing other SOTA baselines, such as ACMGNN and GloGNN. This demonstrates the effectiveness of the proposed \sys.

Despite these positive observations, there are instances on certain datasets where \sys underperforms relative to specific baselines, particularly for node subsets characterized by moderate or mixed homophily levels. In these scenarios, we observe that the gating model may not fully capture the nuanced interaction patterns required for accurate expert selection, which can lead to decreased accuracy for those subsets. In light of these findings, \sys remains highly effective overall, as evidenced by its superior ranking in Table~\ref{table1}. However, the occasional underperformance illustrates the need for further refinements to the gating mechanism, especially with respect to training data diversity and parameter tuning. In future work, augmenting the gating network with structured regularization techniques, developing more adaptive training strategies, or integrating additional expert models could further improve \sys’s ability to handle mixed-homophily scenarios and consistently outperform baseline methods across all graph types.
\begin{figure*}[!ht] 
\centering  
\includegraphics[width=0.6\linewidth]{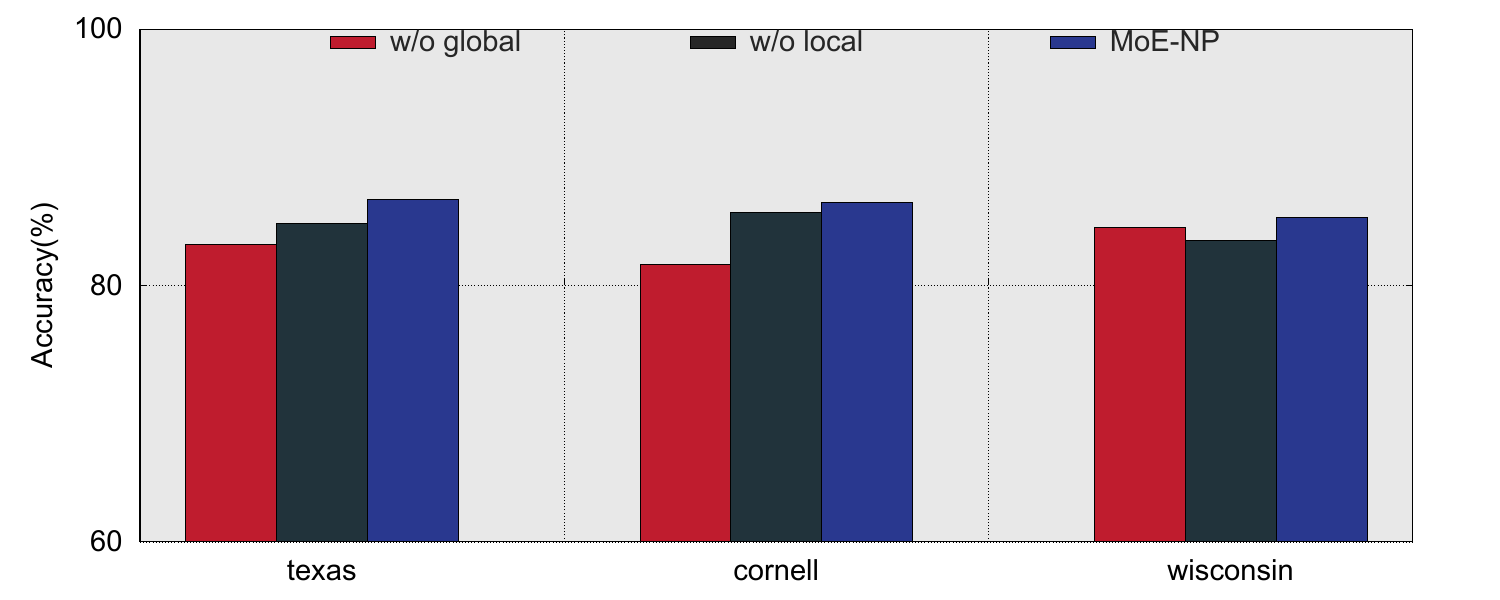}
\caption{Ablation study of \sys on the global or local pattern.}
\label{fig6}  
\end{figure*}

\subsection{Optimization of \sys}
\label{optimization}

The proposed \sys adopts multiple optimization strategies, including (1) an end-to-end approach that simultaneously trains both the gating network and the expert models, and (2) a two-step approach that first trains the experts and subsequently trains the gating network. The following subsections discuss two key elements of these strategies:

\textbf{(1) Load or Train?}  
In a conventional mixture of experts (MoE) framework, pre-trained expert models are typically used directly for subsequent tasks. In our implementation, we first train these experts to leverage domain-specific knowledge, then continue training them within the proposed \sys from their previously states. In this way, experts are expected to learn both the domain knowledge and the transferable node patterns required by \sys.

\textbf{(2) Training Experts with Fewer Samples.}  
Throughout the training process, we observed that if the experts are fully trained on the entire dataset first and then the \sys is trained on the same dataset, the second stage's loss reduction can be minimal. To address this, we train the experts on a smaller subset of the data, thus leaving more room for the gating model to reduce the loss. This revised approach has demonstrated increased effectiveness over the original training procedure.
\begin{table}[!htbp]
\centering
\begin{adjustbox}{width=0.48\textwidth}

\begin{tabular}{|c|c|c|}
\hline \sys& Cora & Chameleon  \\
\hline w/o few samples& $88.45 \pm 1.12$   & $71.82 \pm2.2$   \\  
\hline w few samples& $\mathbf{89.31 \pm 0.91}$   & $\mathbf{73.82 \pm 2.1}$  \\  
\hline
\end{tabular}

\end{adjustbox}
\caption{Comparing the performance of models trained with and without using few-samples training experts}
\label{table3}
\end{table}

\begin{table*}[ht]
\centering 
\begin{adjustbox}{width=\textwidth}  
\begin{tabular}{c c c c c c c c c}
\hline
   & Cora & PubMed & Texas & Cornell & Wisconsin & Chameleon & Actor & \\ \hline
Average & $79.52 \pm 0.81$ & $88.28 \pm 0.49$ & $82.70 \pm 3.67$ & $82.35 \pm 4.64$ & $77.03 \pm 5.83$ & $71.29 \pm 1.69$ & $36.84 \pm 0.87$ &  \\
\textbf{\sys}& $\mathbf{89.31 \pm 0.91}$ & $\mathbf{94.02\pm0.33}$ & $\mathbf{86.76 \pm 4.26}$ & $\mathbf{86.49 \pm 5.54}$ & $\mathbf{85.29 \pm 3.07}$ &$\mathbf{73.82 \pm 2.1}$ & $\mathbf{40.74 \pm 1.36}$   \\
\hline
\end{tabular}
\end{adjustbox}
\caption{Performance comparison of \sys and using average combination.}  
\label{table2}  
\end{table*}

Then we validate the effectiveness of our \sys optimization method by comparing the performance of models trained with and without using few-samples training experts. As shown in Table \ref{table3}, we present the results for the Cora and Chameleon datasets. It is evident that reducing the sample size of the training experts contributes positively to the overall model performance.

\subsection{Analysis of \sys }

\begin{figure*}[!ht]
\centering  
\includegraphics[width=1.5\columnwidth]{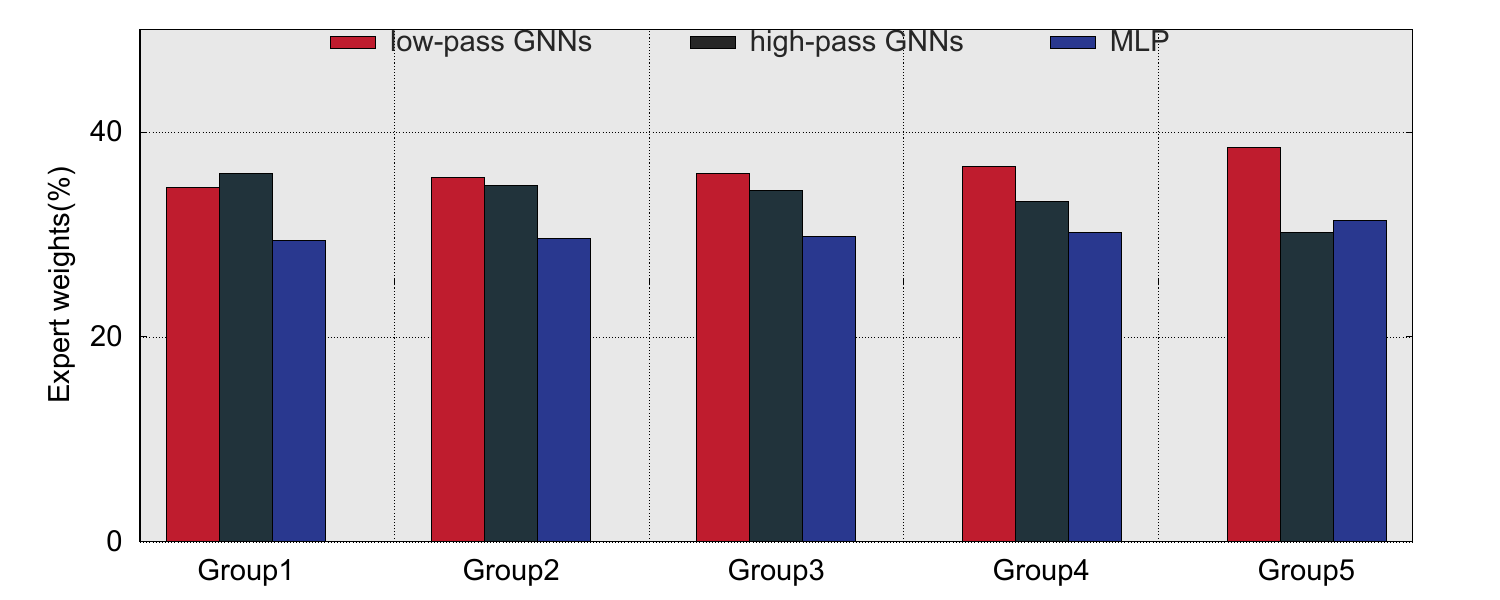}
\caption{The weights assigned to three experts in different groups based on the homophily value, where the group with higher value consists of nodes with higher homophliy values: low-pass GNNs, high-pass GNNs, and MLP, via \sys in chameleon.}

\label{fig4}
\end{figure*}

\begin{figure*}[!ht]
\centering  
\includegraphics[width=1.5\columnwidth]{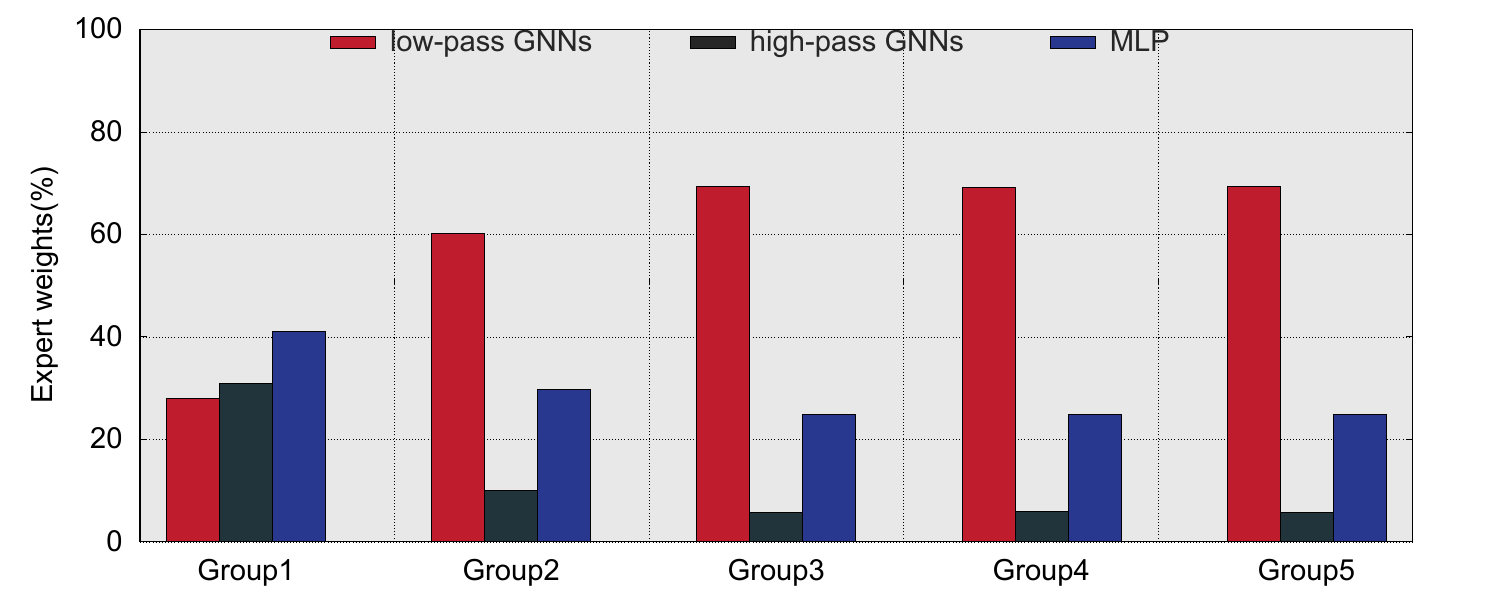}
\caption{The weights assigned to three experts in different groups based on the homophily value, where the group with higher value consists of nodes with higher homophliy values: low-pass GNNs, high-pass GNNs, and MLP, via \sys in PubMed.}

\label{fig7}
\end{figure*}

In this section, we conduct some further analysis towards \sys to demonstrate the rationality and effectiveness of its design. To examine the effectiveness of the gating model and the pattern extractor, we divide nodes in chameleon into five groups, following an increasing order of node homophily value, and then observe the expert weights assigned to different groups in \sys. To simplify the analysis, we implement three basic models as experts. 

As illustrated in Figures \ref{fig4} and \ref{fig7}, the gating model implemented in \sys effectively generates varying expert weights for node groups characterized by different homophily metrics across distinct datasets, namely Chameleon and PubMed. In Figure \ref{fig4}, depicting results from the Chameleon dataset, the pronounced allocation of weights toward low-pass GNNs in the higher homophily groups suggests that these models are particularly well-suited for the classification of densely connected nodes, which aligns with emphical study. Conversely, in groups with lower homophily values, the reliance on high-pass GNNs and MLPs becomes more pronounced, indicating a strategic adaptation to the unique structural features present in those groups.

Similarly, Figure \ref{fig7}, which presents data from the PubMed dataset, reinforces the observations drawn from Chameleon. The expert weight distributions again highlight a preference for low-pass GNNs in the higher homophily groups, while the performance of high-pass GNNs is relatively increased in groups characterized by lower homophily. This consistent finding across different datasets emphasizes the robustness of the gating mechanism in \sys, as it not only adapts the expert weights based on local node relationships but also aligns with broader graph characteristics. These findings collectively affirm the critical role that the gating model plays in optimizing node classification performance through a nuanced understanding of node patterns.

Moreover, we conducted a detailed parameter sensitivity analysis focusing on the maximum length of the random walk, a key factor influencing the amount of node pattern information encoded. Specifically, we evaluated random walk lengths of 5, 10, 20, and 40 on both the Cora and Chameleon datasets. As illustrated in Figure ~\ref{fig5}, the results do not indicate a strictly monotonic relationship wherein longer walks always yield better performance. Instead, the optimal length is dataset-dependent, reflecting the distinct structural characteristics and node feature distributions in each dataset. Furthermore, the analysis suggests that this parameter is not particularly sensitive: moderate deviations from an optimal setting do not lead to substantial performance degradation.

\subsection{Analysis the of experts selection }
We not only use the model proposed in Empirical Study, also add the two node predictors with residual connection. The node predictor with residual means feeds the information that has never been processed from the original input into the next layer, thereby fully exploiting all the information from the initial node features. Simply aggregating representation from the previous layer might lose important information and consequently incur over-smoothing issues. We use a node predictor with residual that can better obtain more informative intermediate representations and avoid the negative influence of neighbors.

\begin{table}[!htbp]
\centering
\begin{adjustbox}{width=0.48\textwidth}
\begin{tabular}{|c|c|c|}
\hline \sys& texas & Chameleon  \\
\hline w/o residual GNNs& $84.9 \pm 5.04$   & $73.05 \pm1.44   $   \\  
\hline w residual GNNs& $\mathbf{86.76 \pm 4.26}$   & $\mathbf{73.82 \pm 2.1}$  \\  
\hline
\end{tabular}
\end{adjustbox}
\caption{Comparing the performance of models trained with and without using fewer samples training experts}
\label{table4}
\end{table}

\vspace{-0.2in}
Then we validate the effectiveness of adding the residual experts of node predictors. We compare the performance of models trained with and without the inclusion of residual experts. As shown in Table \ref{table4}, the results for the Texas and Chameleon datasets are presented. It is evident that adding the residual experts of node predictors positively contributes to the overall model performance.

\subsection{ablation study}
In order to further substantiate the effectiveness of the gating mechanism, we compare \sys\ against an average weight allocation strategy (see Table~\ref{table2}). Furthermore, to validate the rationale behind the node extractor, we perform an ablation study in which we exclude either the global pattern component or the local pattern component. As illustrated in Figure~\ref{fig6}, removing either the global or the local pattern leads to a notable decrease in \sys’s overall performance, underscoring the importance of incorporating both local and global node pattern information.

\section{Related Work}

\sys utilizes a mixture-of-experts paradigm to enhance node classification. In early work, the Graph Convolutional Network (GCN) \cite{kipf2016semi} introduced a novel approach that integrates spectral filtering of graph signals with non-linear transformations for supervised node classification. However, GCNs are known to perform suboptimally on heterophilic graphs, where nodes connected by edges often belong to different classes. Recognizing this limitation, a range of specialized techniques—such as GloGNN \cite{li2022finding}, LinkX \cite{lim2021linkx}, LSGNN \cite{Chen_LSGNN_2023}, Mixhop \cite{abu_Mixhop_2019}, and ACM-GNN \cite{luan_revisiting_2022}—have been proposed to better handle structural variations and improve classification in heterophilic settings. Meanwhile, mounting evidence suggests that real-world graphs commonly exhibit a diverse mixture of node patterns \cite{mao2024demystifying}, reducing the effectiveness of conventional GNNs that assume consistent relationships across all nodes.

Motivated by the complexities revealed in such heterogeneous environments, researchers have turned to the Mixture of Experts (MoE) framework \cite{jacobs1991adaptive,jordan1994hierarchical}, which adopts a divide-and-conquer strategy to assign subsets of tasks to specialized experts. Widely employed in areas like natural language processing \cite{du2022glam,zhou2022mixture} and computer vision \cite{riquelme2021scaling}, MoE techniques have also demonstrated promise in graph-based applications. For instance, GraphMETRO \cite{wu2023graphmetro} leverages MoE to mitigate distribution shifts in GNN models, while Link-MoE \cite{Ma2024MixtureOL} stacks multiple GNNs as experts and adaptively selects the optimal expert for each node pair based on their pairwise attributes. By integrating such principles into \sys, we aim to more effectively accommodate the varied patterns encountered in real-world graphs and ultimately boost node classification performance.

\section{Conclusion}

In this paper, we explored the complex node patterns from different perspectives in the real-world graph datasets and reveal their influences towards node predictors. To accommodate the diverse needs for node classifiers of different nodes, we propose~\sys, a mixture of experts framework for node classification. 
Specifically, \sys combines a mixture of node predictors and  strategically selects models based on node patterns. Extensive experiments demonstrate the proposed \sys demonstrated superior  performance on both homophilic and heterophilic datasets. Further, our theoretical analysis and empirical studies validate the rationality and effectiveness of the proposed \sys.

\begin{acks}
This work was supported by the National Natural Science Foundation of China (project No. 62406336, 62276271), and  the Research Foundation from NUDT (Grant No. ZK2023-12).
\end{acks}

\bibliographystyle{ACM-Reference-Format} 
\balance
\bibliography{ICMR}

\appendix

\section{ Proof of Theorem 2}
\label{proof}
\subsection{Expected Node Features Under CSBM} 

For any node \(i\) :  
\[
\mathbb{E}[\tilde{\mathbf{x}}_i | y_i = 0] = \frac{p \cdot \boldsymbol{\mu} + q \cdot \boldsymbol{\nu}}{p + q} \quad \text{(training)},
\]
\[
\mathbb{E}[\tilde{\mathbf{x}}_i | y_i = 1] = \frac{q \cdot \boldsymbol{\mu} + p \cdot \boldsymbol{\nu}}{p + q} \quad \text{(training)},
\]
\[
\mathbb{E}[\tilde{\mathbf{x}}'_i | y_i = 0] = \frac{p' \cdot \boldsymbol{\mu} + q' \cdot \boldsymbol{\nu}}{p' + q'} \quad \text{(testing)}.
\]
\[
\mathbb{E}[\tilde{\mathbf{x}}'_i | y_i = 0] = \frac{q' \cdot \boldsymbol{\mu} + p' \cdot \boldsymbol{\nu}}{p' + q'} \quad \text{(testing)}.
\]

Follows from CSBM’s definition, where neighbors are sampled with probabilities \(p\) (same class) and \(q\) (different class). The SGC operation \(\mathbf{D}^{-1}\mathbf{A}\mathbf{X}\) averages neighbor features.

\subsection{Classification Margin Analysis}

The optimal classifier \(\mathbf{w}^*\) induces a margin proportional to the separation between class means.  

\textbf{Training Margin:}  
\[
\Delta_{\text{train}} = \mathbf{w}^{*\top} \left(\mathbb{E}[\tilde{\mathbf{x}}|y=1] - \mathbb{E}[\tilde{\mathbf{x}}|y=0]\right) = R \|\boldsymbol{\mu} - \boldsymbol{\nu}\| \cdot \frac{p - q}{p + q}.
\]

\textbf{Test Margin: } 
\[
\Delta_{\text{test}} = R \|\boldsymbol{\mu} - \boldsymbol{\nu}\| \cdot \frac{p' - q'}{p' + q'}.
\]
Substitute \(\mathbf{w}^*\) and class means into the margin definition. Homophily consistency ensures \(\text{sign}(p - q) = \text{sign}(p' - q')\).

\subsection{Cross-Entropy Loss Lower Bound}

\textbf{3.1 Error Probability Lower Bound: }

The projected feature \(z_i | y_i = 1\) follows \(\mathcal{N}(\Delta_{\text{test}}/2, \sigma^2 \|\mathbf{w}^*\|^2)\), and \(z_i | y_i = 0\) follows \(\mathcal{N}(-\Delta_{\text{test}}/2, \sigma^2 \|\mathbf{w}^*\|^2)\).

Misclassification occurs when \(z_i \leq 0\) for \(y_i = 1\) or \(z_i \geq 0\) for \(y_i = 0\). Using the Gaussian tail bound:  
\[
\mathbb{P}(z_i \leq 0 | y_i = 1) = \mathbb{P}\left(Z \geq \frac{\Delta_{\text{test}}}{2\sigma \|\mathbf{w}^*\|}\right),  
\]  
where \(Z \sim \mathcal{N}(0, 1)\).

Applying the inequality \(\mathbb{P}(Z \geq t) \geq \frac{1}{2} \exp(-t^2/2)\) for \(t \geq 0\):  
\[
\mathbb{P}(\hat{y}_i \neq y_i) \geq \frac{1}{2} \exp\left(-\frac{\Delta_{\text{test}}^2}{8\sigma^2}\right).  
\]  
The factor \(8\) arises from \(\|\mathbf{w}^*\| \leq R\) and \(\Delta_{\text{test}} = R \|\boldsymbol{\mu} - \boldsymbol{\nu}\| \cdot \frac{p' - q'}{p' + q'}\).

\textbf{3.2 Cross-Entropy Loss Per Sample}
For misclassified samples, \(\hat{y}_i \leq 0.5\) when \(y_i = 1\), or \(\hat{y}_i \geq 0.5\) when \(y_i = 0\). 

The minimum loss occurs at the decision boundary (\(\hat{y}_i = 0.5\)):  
\[
\ell(0.5, 1) = -\log(0.5) = \log(2),  
\]  
and similarly for \(\ell(0.5, 0)\).  

Since \(\ell(\hat{y}_i, y_i)\) is convex and minimized at \(\hat{y}_i = 0.5\), the inequality holds.
If \(\hat{y}_i \neq y_i\), then:  
\[
\ell(\hat{y}_i, y_i) \geq \log(2).  
\]

\textbf{3.3 Aggregate Loss Lower Bound} 

The expectation of the loss over all test nodes is:  
\[
  L\left(\mathbf{A}^{\prime},\mathbf{X}^{\prime}, \mathbf{w}^*, b^*\right) = \mathbb{E}\left[\ell(\hat{y}_i, y_i)\right].  
\]  

Split the expectation into correctly and incorrectly classified samples:  
\begin{align}
  L\left(\mathbf{A}^{\prime},\mathbf{X}^{\prime}, \mathbf{w}^*, b^*\right) = \mathbb{E}\left[\ell(\hat{y}_i, y_i) | \hat{y}_i = y_i\right] \mathbb{P}(\hat{y}_i = y_i)  \notag\\
   + \mathbb{E}\left[\ell(\hat{y}_i, y_i) | \hat{y}_i \neq y_i\right] \mathbb{P}(\hat{y}_i \neq y_i).  \notag
\end{align}

Dropping the first non-negative term and bounding the second term using Proposition 2, we obtain the following upper bound for the total cross-entropy loss over the test set:
\[
  L\left(\mathbf{A}^{\prime},\mathbf{X}^{\prime}, \mathbf{w}^*, b^*\right) \geq \log(2) \cdot \mathbb{P}(\hat{y}_i \neq y_i).  
\]

\textbf{3.4 Final Lower Bound}
Under the assumptions of Gaussian feature noise and homophily consistency (\((p - q)(p' - q') \geq 0\)), Substitute Proposition 1 into Proposition 3:  
\[
  L\left(\mathbf{A}^{\prime},\mathbf{X}^{\prime}, \mathbf{w}^*, b^*\right) \geq \log(2) \cdot \frac{1}{2} \exp\left(-\frac{\Delta_{\text{test}}^2}{8\sigma^2}\right).  
\]  
Approximate \(\exp(-x) \geq 1 - x\) for \(x \geq 0\) (valid for small \(x\)):  
\[
\exp\left(-\frac{\Delta_{\text{test}}^2}{8\sigma^2}\right) \geq 1 - \frac{\Delta_{\text{test}}}{\sqrt{8}\sigma}.  
\]  
Combine results:  
\[
  L\left(\mathbf{A}^{\prime},\mathbf{X}^{\prime}, \mathbf{w}^*, b^*\right) \geq \log(2) \left(1 - \frac{\Delta_{\text{test}}}{\sqrt{8}\sigma}\right).  
\]  
Substitute \(\Delta_{\text{test}} = R \|\boldsymbol{\mu} - \boldsymbol{\nu}\| \cdot \frac{p' - q'}{p' + q'}\) to obtain the final bound
\[
    L\left(\mathbf{A}^{\prime},\mathbf{X}^{\prime}, \mathbf{w}^*, b^*\right) \geq \log(2) \left(1 - \frac{R \|\boldsymbol{\mu} - \boldsymbol{\nu}\| |p' - q'|}{\sqrt{8}\sigma (p' + q')}\right)(1 + o(1)).  
\] 
\section{Experimental Settings }
For the baseline models, we adopt the same parameter setting in their original paper. For the proposed \sys, we adopt the MLP as the gating model and employ pretrained node predictors as experts. We use a single GPU of NVIDIA RTX 4090 24 GB , to conduct the experiments.

All the hyperparameters are tuned based on the validation set from the following search space:

\textbf{Experts Training}
\begin{itemize}
\item  Learning Rate: $\{0.0005, 0.001\}$
\item Dropout: $\{0 \sim 0.9\}$
\item Weight Decay: $\{1 \mathrm{e}-5,5 \mathrm{e}-5,1 \mathrm{e}-4\}$
\item number of layers = $\{2, 3 ,4\}$
\item hidden channels =$\{32, 64, 128, 256\}$
\end{itemize}

\textbf{Gating Networks Training}

\begin{itemize}
\item Gating Learning Rate: $\{0.0005, 0.001\}$
\item Gating Weight Decay: $\{1 \mathrm{e}-5,5 \mathrm{e}-5,1 \mathrm{e}-4\}$
\item Expert  Learning Rate: $\{0.001,0.01, 0.1,0.5\}$
\item Expert Weight Decay: $\{0,5 \mathrm{e}-5, 5 \mathrm{e}-3 ,5 \mathrm{e}-2\}$
\item hidden channels =$\{32, 64, 128, 256\}$
\item Dropout: $\{0 \sim 0.9\}$
\item number of  MLP layers : $\{1, 2, 3 ,4\}$
\item max random walk length : $\{5,10, 20, 40\}$
\end{itemize}

\end{document}